\documentclass[preprint,12pt]{elsarticle}

\usepackage{amssymb}
\usepackage{booktabs}
\usepackage{multirow}

\journal{Theoretical Computer Science}

\begin{document}

\begin{frontmatter}



\title{Provable Secure Identity Based Generalized Signcryption Scheme}


\author[1]{Gang Yu}\corref{3}
\ead{ygygang@126.com}

\author[2]{Xiaoxiao Ma}
\author[1]{Yong Shen}
\author[1]{Wenbao Han}

\address[1]{Department of Applied Mathematics, Zhengzhou Information Science and Technology Institute, Zhengzhou 450002, China}
\address[2]{School of Surveying and Land Information Engineering of Henan Polytechnic University, Jiaozuo 454003, China}
\cortext[3]{Corresponding author.}

\begin{abstract} According to actual needs, generalized signcryption scheme can flexibly work as an encryption scheme, a
signature scheme or a signcryption scheme. In this paper, firstly,
we give a security model for identity based generalized signcryption
which is more complete than existing model. Secondly, we propose an
identity based generalized signcryption scheme. Thirdly, we give the
security proof of the new scheme in this complete model. Comparing
with existing identity based generalized signcryption, the new
scheme has less implementation complexity. Moreover, the new scheme
has comparable computation complexity with the existing normal
signcryption schemes.
\end{abstract}


\end{frontmatter}


\section{Introduction}

Encryption and signature are fundamental tools of Public Key
Cryptography for confidentiality and authenticity respectively.
Traditionally, these two main building-blocks have been considered
as independent entities. However, these two basic cryptographic
techniques may be combined together in various ways, such as
sign-then-encrypt and encrypt-then-sign, in many applications to
ensure privacy and authenticity simultaneously. To enhance
efficiency, Zheng \cite{ZHE} proposed a novel conception named
signcryption, which can fulfill both the functions of signature and
encryption in a logical step. Comparing with the traditional
methods, signcryption has less computation complexity, less
communication complexity and less implementation complexity. Just
because signcryption scheme has so many advantages and extensive
application prospective, many public key based signcryption schemes
have been proposed \cite{ZHI, BD, JCL,HLS}.

Identity-based cryptography was introduced by Shamir \cite{SHA} in
1984, in which the public keys of users are respectively their
identities and the secret keys of users are created by a credit
third party named Public Key Generator (PKG). In this way, the
identity-based cryptography greatly relieves the burden of public
key management and provides a more convenient alternative to
conventional public key infrastructure. In \cite{SHA}, Shamir
proposed an identity based signature scheme but for many years there
wasn't an identity based encryption scheme. Until 2001, Boneh and
Franklin\cite{BF} using bilinear pairing gave a practical secure
identity based encryption scheme. The first identity based
signcryption scheme was proposed by Malone-Lee \cite{ML} along with
a security model. Since then, many identity based signcryption
schemes are proposed \cite{LQ, BO, CYH, CM}.

Signcryption has considered these application environments that need
simultaneous message privacy and data integrity. However, in some
applications, these two properties are not essential. That is,
sometimes only message confidentiality is needed or sometimes only
authenticity is needed. In this case, in order to ensure privacy or
authenticity separately, signcryption must preserve sign module or
encryption module, which must increase the corresponding computation
complexity and implementation complexity. To decrease implementation
complexity, Han et al. \cite{HY} proposed a new primitive called
generalized signcryption, which can work as an encryption scheme, a
signature scheme or a signcryption scheme, and gave an generalized
signcryption based on ECDSA. Wang et al. \cite{WYH} gave the formal
security notions for this new primitive and improved the original
generalized signcryption proposed by Han et al. \cite{HY}. In
\cite{WYH}, Wang et al. pointed out some open problems. One of
theses problems is to enhance efficiency. Another of theses problems
is to design identity based generalized signcryption scheme.

Lal et al. \cite{LK} gave an identity based generalized signcryption
scheme (IDGSC). However, after much study, we find his security
model is not complete. And his scheme is not secure under the
complete security model for IDGSC. In this paper, our main works
include three aspects. Firstly, in the second section, we give the
definition of IDGSC and the security model for IDGSC. Secondly, in
Section 3, we propose an efficient IDGSC. Thirdly, in Section 4, we
give the efficiency analysis and security results.

\section{IDGSC and Its Security
Notions}

\subsection{Definition of IDGSC}

Firstly, we will review the algorithm constitution of identity based
encryption (IDEC), identity based signature (IDSG) and identity
based signcryption (IDSC). Then, we will introduce the algorithms
that consist of an identity based generalized signcryption (IDGSC).

\noindent \textbf{Definition 1.} A normal identity based encryption
scheme \[ IDEC = (Setup,Extract,Encrypt,Decrypt)
\]
consists of four algorithms.

\noindent \textbf{Setup:} This is the system initialization
algorithm. On input of the security parameter $1^k $, this algorithm
generates the system parameters $params$ and the PKG generates his
master key $s$ and public key $P_{Pub} $. The global public
parameters include $params$ and $P_{Pub} $. We write
$((params,P_{Pub} ),s) \leftarrow Setup(1^k )$.

\noindent \textbf{Extract:} This is the user key generation
algorithm. Given some user's identity $ID$, PKG uses it to produce a
pair of corresponding public/private keys. We write $(S_{ID} ,Q_{ID}
) \leftarrow Extract(ID,s)$.

\noindent \textbf{Encrypt:} It takes as input a receiver's identity
$ID_r$ and a message $m$, using the public parameters
$(params,P_{Pub} )$, outputs a ciphertext $\varepsilon$. We write
$\varepsilon \leftarrow Encrypt(ID_r ,m)$.

\noindent\textbf{Decrypt:} It takes as input a receiver's private
key $ S_r $ and a ciphertext $\varepsilon$, using the public
parameters $(params,P_{Pub} )$, outputs a message $m$ or the invalid
symbol ${ \bot}$. We write $m \leftarrow Decrypt(S_r ,\varepsilon
)$.

\noindent\textbf{Definition 2. } A normal identity based signature
scheme \[ IDSG = (Setup,Extract,Sign,Verify)
\]
consists of four algorithms.

\noindent \textbf{Setup:} It is the same as the corresponding Setup
algorithm in Definition 1.

\noindent \textbf{Extract:} It is the same as the Extract algorithm
in Definition 1.

\noindent \textbf{Sign:} This algorithm takes as input a signer's
private key $ S_s $ and a message $m$, using the public parameters
$(params,P_{Pub} )$, outputs a signature $ \sigma $. We write
$\sigma \leftarrow Sign(S_s ,m)$.

\noindent \textbf{Verify:} This algorithm takes as input the
signer's public key $ Q_s $, a message $m$ and the corresponding
signature $ \sigma $, and outputs the valid symbol $\top$ or the
invalid symbol $\bot$. We write $(\top\,or\,\bot) \leftarrow
Verify(Q_s ,m,\sigma )$.

\noindent\textbf{Definition 3. } A normal identity based
signcryption scheme \[ IDSC = (Setup,Extract,Signcrypt,Unsigncrypt)
\]
consists of four algorithms.

\noindent \textbf{Setup:} It is the same as the corresponding Setup
algorithm in Definition 1.

\noindent \textbf{Extract:} It is the same as the Extract algorithm
in Definition 1.

\noindent \textbf{Signcrypt:} This algorithm takes as input the
sender's private key $ S_s $, the receiver's public key $ Q_r $ and
a message $m$, using the public parameters $(params,P_{Pub} )$,
outputs a ciphertext $\delta $. We write $\delta \leftarrow SC(S_s
,Q_r ,m)$.

\noindent \textbf{Unsigncrypt:} This algorithm takes as input the
sender's public key $ Q_s $, the receiver's secret key $ S_r $ and a
ciphertext $\delta$, using the public parameters $(params,P_{Pub}
)$, outputs a message $m$ or the invalid symbol ${ \bot}$. We write
$m \leftarrow UC(Q_s ,S_r ,\delta )$.

Generalized signcryption scheme can work as encryption scheme,
signature scheme and signcryption scheme according to different
needs. Let $ IDSG = (Setup,Extract,Sign,Verify) $, $ IDEC =
(Setup,Extract,Encryp\\t,Decrypt) $ and $ IDSC =
(Setup,Extract,Signcrypt,Unsigncrypt) $ respectively be an identity
based signature scheme, encryption scheme and signcryption scheme.

\noindent\textbf{Definition 4. }An identity based generalized
signcryption scheme $IDGSC = (Setup,Extract,GSC,GUC)$consists of
following four algorithms:

\noindent \textbf{Setup:} It is the same as the corresponding Setup
algorithm in Definition 1.

\noindent \textbf{Extract:} It is the same as the Extract algorithm
in Definition 1.

\noindent\textbf{GSC: } for a message $m$,

\noindent-When $ID_s  \in \Phi (ID_s  = 0)$, $\varepsilon \leftarrow
GSC(\Phi ,Q_r ,m) = Encrypt(Q_r ,m)$.

\noindent-When $ID_r  \in \Phi (ID_r  = 0)$, $ \sigma  \leftarrow
GSC(S_s ,\Phi ,m) = Sign(S_s ,m)$.

\noindent-When $ID_s  \notin \Phi ,ID_r  \notin \Phi $, $\delta
\leftarrow GSC(S_s ,Q_r ,m) = SC(S_s ,Q_r ,m)$.

\noindent\textbf{GUC: } to unsigncryt a ciphertext $\delta$,

\noindent-When $ID_s  \in \Phi (ID_s  = 0)$, $m \leftarrow GUC(\Phi
,Q_r ,\varepsilon ) = Decrypt(Q_r ,\varepsilon )$.

\noindent-When $ID_r  \in \Phi (ID_r  = 0)$, $ (\top,\bot)
\leftarrow GUC(S_s ,\Phi ,\sigma ) = Verify(S_s ,\sigma )$.

\noindent-When $ID_s  \notin \Phi ,ID_r  \notin \Phi $, $m
\leftarrow GUC(Q_s ,S_r ,\delta ) = UC(Q_s ,S_r ,\delta ) $.

\subsection{Security models for IDGSC}

 In our security
model, there are seven types of queries that the adversary $A$ may
inquire the challenger $C$ for answers. In the following text,
``$Alice\{Text1\}\rightarrow Bob$, and then $Bob\{Text2\}\rightarrow
Alice$" denotes that Alice submits Text1 to Bob, and then Bob
responds with Text2 to Alice.

\noindent Extract query: $A\{ID\}\rightarrow C$, and then $C\{
{S_{ID}  = Extract(ID)} \}\rightarrow A$

\noindent Sign query: $A\{ {ID_s ,m} \}\rightarrow C$, and then $C\{
{\sigma = Sign(S_s ,m)}
 \}\rightarrow A$

\noindent Verify query: $A\{ {ID_s ,\sigma }
 \}\rightarrow C$, and then $C\{
{(\top\,or\,\bot) = Verify(Q_s ,\sigma )}
 \}\rightarrow A$

\noindent Encrypt query: $A\{ {ID_r ,m} \}\rightarrow C$, and then
$C\{ {\varepsilon  = Encrypt(Q_r ,m)}
 \}\rightarrow A$

\noindent Decrypt query: $A\{  {ID_r ,\varepsilon }  \}\rightarrow
C$, and then $C\{{m = Decrypt(S_r ,\varepsilon )}
 \}\rightarrow A$

\noindent GSC query: $A\{ {ID_s ,ID_r ,m} \}\rightarrow C$, and then
$C\{ {\delta = GSC(S_s ,Q_r ,m)}
 \}\rightarrow A$

\noindent GUC query: $A\{ {ID_s ,ID_r ,\delta } \}\rightarrow C$,
and then $C\{{m = GUC(Q_s ,S_r ,\delta )}
 \}\rightarrow A$

The generalized signcryption can work in three modes: in signature
mode, in encryption mode and in signcryption mode, denoted
IDGSC-IN-SG, IDGSC-IN-EN and IDGSC-IN-SC respectively. Firstly, we
define the confidentiality of IDGSC-IN-EN (Def. 5) and IDGSC-IN-SC
(Def. 6) separately.

\noindent\textbf{Definition 5. } IND-(IDGSC-IN-EN)-CCA Security

Consider the following game played by a challenger $C$ and an
adversary $A$.

\textbf{Game 1 }

\noindent\emph{Initialize.} Challenger $C$ runs  $Setup(1^k )$ and
sends the public parameters $(params,P_{Pub} )$
 to the adversary $A$. $C$ keeps master key $s$ secret.

\noindent\emph{Phase 1.} In Phase 1, $A$ performs a polynomially
bounded number of above seven types of queries. These queries made
by $A$ are adaptive; that is every query may depend on the answers
to previous queries.

\noindent\emph{Challenge.} The adversary $A$ chooses two identities
$ID_A  = 0,ID_B\neq0 $ and two messages $m_0 ,m_1 $. Here, the
adversary $A$ cannot have asked Extract query on $ID_B $ in Phase 1.
The challenger $C$ flips a fair binary coin $\gamma $, encrypts
$m_\gamma  $ and then sends the target ciphertext $\varepsilon ^* $
to $A$.

\noindent\emph{Phase 2.} In this phase,  $A$ asks again a
polynomially bounded number of above queries just with a natural
restriction that he cannot make Extract queries on $ID_B $, and he
cannot ask Decrypt query on target ciphertext $\varepsilon ^* $.

\noindent\emph{Guess.} Finally,  $A$ produces his guess $ \gamma
^{'}$ on $\gamma $, and wins the game if ${\gamma ^{'}=\gamma }$.

$A$'s advantage of winning Game 1 is defined to be $Adv_{A^{idgsc -
in - en} }^{ind - cca2} (t,p) = |2P[\gamma ^{'}  = \gamma ] - 1|$.
We say that identity based generalized signcryption in encryption
mode is IND-(IDGSC-IN-EN)-CCA secure if no polynomially bounded
adversary $A$ has a non-negligible advantage in Game 1.

\noindent\textbf{Definition 6.} IND-(IDGSC-IN-SC)-CCA Security

Consider the following game played by a challenger $C$ and an
adversary $A$.

\textbf{Game 2 }

\noindent\emph{Initialize.} and \noindent\emph{Phase 1.}

Challenger $C$ and adversary $A$ act the same as what they do in the
corresponding stage in Game 1.

\noindent\emph{Challenge.} The adversary $A$ chooses two identities
$ID_A\neq0,ID_B\neq0 $ and two messages $m_0 ,m_1 $. Here, the
adversary $A$ cannot have asked Extract query on $ID_B $ in Phase 1.
The challenger $C$ flips a fair binary coin $\gamma $, signcrypts
$m_\gamma  $ and then sends the target ciphertext $\delta ^* $ to
$A$.

\noindent\emph{Phase 2.} In this phase,  $A$ asks a polynomially
bounded number of above queries just with a natural restriction that
he cannot make Extract queries on $ID_B $, and he cannot ask
Unsigncrypt query on target ciphertext $\delta ^* $.

\noindent\emph{Guess.} Finally,  $A$ produces his guess $ \gamma
^{'}$ on $\gamma $, and wins the game if ${\gamma ^{'}=\gamma }$.

$A$'s advantage of winning Game 1 is defined to be$ Adv_{A^{idgsc -
in - sc} }^{ind - cca2} (t,p) = |2P[\gamma ^{'}  = \gamma ] - 1|$.
We say that identity based generalized signcryption in signcryption
mode is IND-(IDGSC-IN-SC)-CCA secure if no polynomially bounded
adversary $A$ has a non-negligible advantage in Game 2.

\textbf{Note 1.} The differences between Def. 5 and Def. 6 deserve
to be mentioned. Firstly, in Phase 2 of Def. 5, the adversary is
prohibited from making Decrypt query on the challenge ciphertext.
However, he can transform the challenge ciphertext into some valid
signcryption ciphertext and make Unsigncrypt query on the
corresponding signcryption ciphertext. Secondly, the adversary is
restricted not to make Unsigncrypt query on the challenge ciphertext
in Phase 2 of Def. 6. But, he can transform the challenge ciphertext
into some valid encryption ciphertext and make Decrypt query on the
corresponding encryption ciphertext. Such differences are not
considered in the security model proposed by S. Lal et al.
\cite{LK}.

Secondly, we define the unforgeability of IDGSC-IN-SG (Def.7) and
IDGSC-IN-SC (Def.8) separately.

\noindent\textbf{Definition 7.} EF-(IDGSC-IN-SG)-ACMA Security

Consider the following game played by a challenger $C$ and an
adversary $A$.

\textbf{Game 3 }

\noindent\emph{Initialize.} Challenger $C$ runs  $Setup(1^k )$ and
sends the public parameters $(params,P_{Pub} )$
 to the adversary $A$. $C$ keeps the master key $s$ secret.

\noindent\emph{Probe.} In this phase,  $A$ performs a polynomially
bounded number of above seven kinds of queries.

\noindent\emph{Forge.} Finally, $A$ produces two identities $ID_A
,ID_B $, where $ID_B  = 0$, and a ciphertext $\sigma ^*  = (X^* ,m^*
,V^* )$. The adversary wins the game if: $ID_A  \ne 0$; $Verify(m^*
,ID_A ,(X^* ,V^* )) = \top$; no Extraction query was made on $ID_A
$; $(X^* ,V^* )$  was not result from $Sign(m^* )$ query with signer
$ID_A$.

We define the advantage of $A$ to be $ adv_{A^{idgsc - in - sg}
}^{ef - acma} (t,p) = \Pr [Awins]$. We say that an identity based
generalized signcryption in signature mode is EF-(IDGSC-IN-SG)-ACMA
secure if no polynomially bounded adversary has a non-negligible
advantage in Game 3.

\noindent\textbf{Definition 8.} EF-(IDGSC-IN-SC)-ACMA Security

Consider the following game played by a challenger $C$ and an
adversary $A$.

\textbf{Game 4 }

\noindent\emph{Initialize.} Challenger $C$ runs  $Setup(1^k )$ and
sends the public parameters $(params,P_{Pub} )$
 to the adversary $A$. $C$ keeps the master key $s$ secret.

\noindent\emph{Probe.} In this phase,  $A$ performs a polynomially
bounded number of above seven kinds of queries.

\noindent\emph{Forge.} Finally, $A$ produces two identities $ID_A
,ID_B $, and a ciphertext $\sigma ^*  = (X^* ,C^* ,V^* )$. Let $m^*
$ be the result of unsigncrypting $\delta ^* $ under the secret key
corresponding to $ID_B $.The adversary wins the game if: $ID_A  \ne
0$; $ID_A  \ne ID_B $; $Verify(m^* ,ID_A ,(X^* ,V^* )) = \top$; no
Extraction query was made on $ID_A$; $(\delta ^*,ID_A,ID_B) $ wasn't
outputs by a Signcrypt query.

We define the advantage of $A$ to be $Adv_{A^{idgsc - insc} }^{ef -
acma} (t,p) = \Pr [Awins]$. We say that an identity based
signcryption in signcryption mode is EF-(IDGSC-IN-SC)-ACMA secure if
no polynomially bounded adversary has a non-negligible advantage in
Game 4.

\textbf{Note 2.} The differences between Def. 7 and Def. 8 also need
to be noticed. In Def. 7, the forged signature is not obtained from
the Sign query. But it can be transformed from some valid
signcryption ciphertext that is gotten from Signcrypt query. In
contrast, in Def. 8, the forged signcryption cipherxt is not the
output of Signcrypt query. But it can be transformed from some
answer of the Sign query. Such differences are not considered in the
security model proposed by S. Lal et al. \cite{LK}. Consequently, in
S. Lal et al. \cite{LK}'s scheme, adversary can easily forge a valid
signature through a correspondingly Signcrypt query and Unsigncrypt
query.

\section{Our Scheme}

\subsection{Description of our scheme}

Before describing our scheme we need to define a special function $
f(ID)$, where $ ID \in \{ 0,1\} ^{n_1 }$. If identity is vacant,
that is $ID \in \Phi$, let $ID = 0$, $f(ID) = 0$; in other cases,
$f(ID) = 0$. The concrete algorithms of our scheme are described as
follows.

\noindent\textbf{Setup:} Given the security parameter $1^k$, this
algorithm outputs: two cycle groups $(G_1 , + )$ and $(G_2 , \cdot
)$ of prime order $q$, a generator $P$ of $G_1$, a bilinear map $
\hat e:G_1  \times G_1  \to G_2$ between $G_1$ and $G_2$, four hash
functions:

$H_0 :\{ 0,1\} ^{n_1 }  \to G_1 ^*$\ ;\ \ \ \     $ H_1 :G_2  \to \{
0,1\} ^{n_2 }\times\{ 0,1\} ^{n_1 }\times G_{1} ^{*}$\ ;

$H_2 :\{ 0,1\} ^{n_2 }\times\{ 0,1\} ^{n_1 }\times\{ 0,1\} ^{n_1 }
\to Z_q^*$\ ; \ \ \ \   $H_3 :\{ 0,1\} ^{n_2 }  \times G_1  \to
Z_q^*$\ .

\noindent Where $n_1$ and $n_2$ respectively denote the bit length
of user's identity and the message.  Here $H_0,H_1$ needs to satisfy
an additional property: $H_{0}(0)=\vartheta,H_{1}(1)=0$, where
$\vartheta$ denotes the infinite element in group $G_1$.The system
parameters are $params = \{ G_1 ,G_2 ,q,n_1 ,n_2 ,\hat e,P,H_0 ,H_1
,$ $H_2 ,H_3 \}$. Then, PKG chooses $s$ randomly from $Z_q^*$ as his
master key, and computes $ P_{Pub} = sP$ as his public key. The
global public parameters are $ (params,P_{Pub} ) = \{ G_1 ,G_2
,q,n_1 ,n_2 ,\hat e,P,P_{Pub} ,H_0 ,H_1 ,H_2 ,H_3 \}$.

\noindent\textbf{Extract:} each user in the system with identity
$ID_U$, his public key $ Q_U  = H_0 (ID_U )$ is a simple transform
from his identity. Then PKG computes private key $ S_U  = sQ_U$ for
$ID_U$.

\noindent\textbf{Generalized Signcryption:} Suppose Alice with
identity $ID_A$ wants to send message $m$ to Bob whose identity is
$ID_B$, he does as following:

\noindent - Computes $ f(ID_A ) $ and $f(ID_B ) $.

\noindent - Selects $r$ uniformly from $Z_q^* $, and computes $ X =
rP $.

\noindent - Computes $ h_2  = H_2 (m||ID_A||ID_B) $ and $ h_3  = H_3
(m||X)$.

\noindent - Computes $ V = r^{ - 1} (h_2P + f(ID_A ) \cdot h_3 \cdot
S_A )$.

\noindent - Computes $ Q_B  = H_0 (ID_B )$ and $w = \hat e(P_{Pub}
,Q_B )^{r \cdot f(ID_B )}$.

\noindent - Computes $h_1  = H_1 (w) $ and $ y = m||ID_A||V \oplus
h_1 $.

\noindent - Sends $ (X,y)$ to Bob.

\noindent\textbf{Generalized Unsigncryption:} After receiving $
(X,y)$:

\noindent - Computes $f(ID_B ) $.

\noindent - Computes $ w = \hat e(X,S_B )^{f(ID_B )}$, $ h_1 = H_1
(w)$, $m||ID_A||V = y \oplus h_1$.

\noindent - Computes $ h_2  = H_2 (m||ID_A||ID_B) $  and $ h_3  =
H_3 (m||X)$.

\noindent - Checks that $ \hat e(X,V) = \hat e(P,P)^{h_2 } \cdot
\hat e(P_{Pub} ,Q_A )^{h_3  \cdot f(ID_A )}$, if not, returns$\bot$.
Else, returns $m$.

\subsection{Correctness}

There are three cases to be considered.

\textbf{Case 1.} IDGSC-IN-SC

In this case, there is $ID_A ,ID_B  \notin \Phi $ (That is $ID_A
,ID_B \ne 0$), so $ f(ID_A ) = f(ID_B ) = 1 $ and the scheme is
actually a signcryption scheme. It is easy to verify that:

\noindent $ w = \hat e(P_{Pub} ,Q_B )^{r }  = \hat e(X,S_B ) $;

\noindent $ \hat e(X,V) = \hat e(rP,r^{ - 1} (h_2 P + h_3 \cdot S_A
)= \hat e(P,P)^{h_2 }  \cdot \hat e(P_{Pub} ,Q_A )^{h_3} $;

\noindent $ UC(ID_A ,ID_B,SC(ID_A ,ID_B,m ) ) = m$.

\noindent So our scheme in signcryption mode is correct.

\textbf{Case 2.} IDGSC-IN-SG

In this case, there is $ID_A \notin \Phi,ID_B  \in \Phi $ (That is
$ID_A \ne 0 ,ID_B=0$.), so $ f(ID_A ) = 1,f(ID_B ) = 0 $.  The
generalized signcryption scheme in signature mode is as follows:

\noindent\textbf{Sign:}

\noindent - Selects $r$ uniformly from $Z_q^* $, and computes $ X =
rP $.

\noindent - Computes $ h_2  = H_2 (m||ID_A||0) $ and $ h_3  = H_3
(m||X)$.

\noindent - Computes $ V = r^{ - 1} (h_2P + f(ID_A ) \cdot h_3 \cdot
S_A )= r^{ - 1} (h_2P + h_3 \cdot S_A )$.

\noindent - Computes $ Q_B  = H_0 (0 )=\vartheta$ and $w = \hat
e(P_{Pub} ,\vartheta )^{r \cdot {f(ID_B )}}=1$.

\noindent - Computes $h_1  = H_1 (w)= H_1 (1)=0 $ and $ y =
m||ID_A||V \oplus 0= m||ID_A||V$.

\noindent - Outputs the signature$ (X,m||ID_A||V)$.

\noindent\textbf{Verify:}

\noindent - Computes $ h_2  = H_2 (m||ID_A||0) $ and $ h_3  = H_3
(m||X)$.

\noindent - Checks that $ \hat e(X,V) = \hat e(P,P)^{h_2 } \cdot
\hat e(P_{Pub} ,Q_A )^{h_3}$, if not, returns$\bot$.

In fact, the reduced signature scheme is the signature scheme,
denoted PSG, proposed by Paterson \cite{P}.

\noindent\textbf{Case 3.} IDGSC-IN-EN

In this case, there is $ID_A \in \Phi,ID_B  \notin \Phi $ (That is
$ID_A=0 ,ID_B \ne 0$.), so $ f(ID_A ) = 0,f(ID_B ) = 1 $. The
generalized signcryption scheme in encryption mode is as follows:

\noindent\textbf{Encrypt:}

\noindent - Selects $r$ uniformly from $Z_q^* $, and computes $ X =
rP $.

\noindent - Computes $ h_2  = H_2 (m||0||ID_B) $ and $ h_3  = H_3
(m||X)$.

\noindent - Computes $ V = r^{ - 1} (h_2P + f(ID_A ) \cdot h_3 \cdot
S_A )= r^{ - 1} h_2P$.

\noindent - Computes $ Q_B  = H_0 (ID_B )$ and $w = \hat e(P_{Pub}
,Q_B )^{r }$.

\noindent - Computes $h_1  = H_1 (w) $ and $ y = m||0||V \oplus h_1
$.

\noindent - Sends $ (X,y)$ to Bob.

\noindent\textbf{Decrypt:}

\noindent - Computes $f(ID_B ) $.

\noindent - Computes $ w = \hat e(X,S_B )^{f(ID_B )}= \hat e(X,S_B
)$ and $ h_1 = H_1 (w)$.

\noindent - Computes $m||0||V = y \oplus h_1$.

\noindent - Computes $ h_2  = H_2 (m||0||ID_B) $ and $ h_3  = H_3
(m||X)$.

\noindent - Checks that $ \hat e(X,V) = \hat e(P,P)^{h_2 }$. if not,
returns$\bot$. Else, returns $m$.

Actually, the reduced encryption scheme is combination of the basic
encryption scheme, denoted BFE, proposed by Boneh and Franklin
\cite{BF} and a one-time signature scheme.

\section{ Efficiency Analysis and Security Results}

\subsection{Efficiency Analysis}

The main purpose of generalized signcryption is to reduce
implementation complexity. According to different application
environments, generalized signcryption can fulfill the function of
signature, encryption or signcryption respectively. However, the
computation complexity may increase comparing with normal
signcryption scheme. Such as, \cite{HY, WYH}, these schemes all need
an additional  secure MAC function which not only increase the
computation complexity but also the implementation complexity.
Fortunately, this additional requirements are not needed in our
scheme. Moreover, our scheme is as efficient as \cite{CM}, which is
the most efficient identity based signcryption scheme. In Table 1
below we compare the computation complexity of our scheme, denoted
NIDGSC, with several famous signcryption schemes. We use mul., exps.
and cps. as abbreviations for multiplications, exponentiations and
computations respectively. Here, the computations that can be
pre-calculated will be denoted by $(+?)$.

\begin{center}

\begin{tabular}{|c|c|c|c|c|c|c|}
  \hline
 \multirow {3}{*}{Schemes} & \multicolumn{3}{|c|}{Sign/Encrypt} & \multicolumn{3}{|c|}{Decrypt/Verify} \\
 \cline{2-7}
    & mul. & exps.  & $\hat{e}$ & mul.  & exps.
 & $\hat{e}$\\
 & in $G_1$ & in $G_2$ & cps. & in $G_1$ & in $G_2$ & cps.\\
\hline
  \cite{ML} & 3 & 0 & 0(+1) & 0 & 1 & 3(+1) \\
  \cite{LQ} & 2 & 2 & 0(+2) & 0 & 2 & 2(+2) \\
  \cite{BO} & 3 & 1 & 0(+1) & 2 & 0 & 3(+1) \\
  \cite{CYH} & 2 & 0 & 0(+2) & 1 & 0 & 4 \\
  \cite{CM} & 3 & 0 & 0(+1) & 1 & 0 & 3 \\
  \cite{LK} & 5 & 0 & 0(+1) & 1 & 0 & 3(+1) \\
  NIDGSC & 3 & 1 & 0(+1) & 0 & 2 & 2(+2) \\
  \hline
\end{tabular}
\end{center}

\noindent\textbf{Table 1.} Comparasion between the dominant
operations required for IDGSC and other schemes

\subsection{Security Results}

In this section we will state the security results for our scheme
under the security model defined in Section 2.2. Our results are all
in the random oracle model. In each of the results below we assume
that the adversary makes $ q_i$ queries to $ H_i$ for $ i =
0,1,2,3$. $ q_s$ and $ q_u$ denote the number of Signcrypt and
Unsigncrypt queries made by the adversary respectively. $ n_3$ and $
n_4$ denote the bit length of an element in group $G_1$ and $G_2$
respectively.

\textbf{Theorem 1.} If there is an EF-ACMA adversary $A$ of NIDGSC
in signature-mode that succeeds with advantage $adv_{A^{idgsc - in -
sg} }^{ef - acma} (t,p)$, then there is a simulator $C$ that can
forge valid signature of PSG with advantage $ \xi \approx
adv_{A^{idgsc - in - sg} }^{ef - acma} (t,p)$.

When NIDGSC works as a signature scheme, it is actually the
signature scheme, PSG, proposed by Paterson \cite{P}. The PSG scheme
itself is EF-ACMA secure. Considering Signcrypt/Unsigncrypt query
that is absent in normal signature scheme, these queries are useless
to the adversary of NIDGSC-IN-SG. Because the identities of sender
and receiver are included in the signature. There are two ways to
modify these values. First, the adversary must to find a special
Hash collision. Second, the adversary succeeds in solving the ECDLP
\cite{CR} problem. In such cases, the adversary has negligible
advantage to modify these values. So an EF-ACMA adversary can attack
PSG scheme if he can attack NIDGSC in signature mode.

\textbf{Theorem 2.} Let $Adv_{A^{idgsc - in - en} }^{ind - cca2}
(t,p)=\xi$ be advantage of an IND-CCA2 adversary $A$ of NIDGSC in
encryption-mode, then $\xi$ is polynomial time negligible.

When NIDGSC works as an encryption scheme, it is actually the
combination of the basic identity based encryption scheme proposed
by \cite{BF} and a one-time signature scheme. Owing to the theorem
proposed by Canetti et al.\cite{CHK}, this combined encryption
scheme is secure against normal adaptive chosen-ciphertext attack.
Considering Signcrypt/Unsigncrypt query, the adversary can not
transform the target encryption ciphertext into a valid signcryption
ciphertext. This conclusion is based on the EF-ACMA security of PSG.
So NIDGSC in encryption mode is IND-CCA2 secure.

\textbf{Theorem 3.} If $A$ can forge valid signcryption ciphertext
of NIDGSC in signcryption-mode successfully with advantage
$Adv_{A^{idgsc - insc} }^{ef - acma} (t,p)$, then there is a
simulator $C$ that can forge valid signature of PSG with advantage
$\xi$:

$ \xi  \geqslant Adv_{A^{idgsc - insc} }^{ef - acma} (t,p) + {(q_1
\cdot q_s )} /{2^{n_4 }}  + {q_u } /{({2^{n_2 } }\cdot{2^{n_1 }
}\cdot{2^{n_3 } }) } $.

The corresponding proofs are given in Appendix A.

\textbf{Theorem 4. }  If there is an IND-IBSC-CCA2 adversary $A$ of
NIDGSC in signcryption-mode that succeeds with advantage
$Adv_{A^{idgsc - in - sc} }^{ind - cca2} (t,p)$, then there is a
challenger $C$ running in polynomial time that solves the weak BCDH
problem with advantage $\xi$:

$ \xi \geqslant Adv_{A^{idgsc - in - sc} }^{ind - cca2} (t,p)/{(q_{0
\cdot } q_1 )}$.

The definition of weak BCDH problem and corresponding proofs are
given in Appendix B.

\section{Conclusions}

In this paper, we define the security model for IDGSC and propose an
efficient IDGSC which is proved secure under this security model.
Comparing with existing generalized signcryption schemes, our scheme
doesn't need an extra secure MAC function. So it has less
implementation complexity. What's more, it is almost as efficient as
the normal signcryption scheme.

An interesting open question is to design a non-ID based (public key
or Certificateless) generalized signcryption scheme that does not
need an additional MAC function.

\begin{flushleft}

{\bf  Acknowledgement}

\end{flushleft}

This work is supported by 863 Project of China (No. 2009AA01Z417).
The authors would like to thanks the anonymous referees for their
helpful comments.

\begin{flushleft}

{\bf  References}

\end{flushleft}



\bibliographystyle{elsarticle-num}
\bibliography{<your-bib-database>}





\appendix

\section{Proof of Theorem 3}

We will reduce the attack to EF-ACMA of NIDGSC to EF-ACMA of PSG
proposed by Paterson \cite{P}. Hence, we define two experiment Exp 1
and Exp 2. In each experiment, the private and public key and the
Random Oracle's coin flipping space are not changed. The difference
between Exp 1 and Exp 2 comes from rules of oracle service that
challenger provides for the adversary.

\noindent\textbf{Exp 1}

In this experiment, we use the standard technique to simulate Hash
functions used in our scheme. It is well-known that no adversary can
distinguish between this environment and the real environment in
polynomially bounded time. Let $S_0$ denote the event that EF-ACMA
adversary can attack NIDGSC successfully in Exp 1.

Challenger $C$ needs to keep four lists $L_i ,i = 0,1,2,3 $ which
are vacant at the very beginning. These lists are used to record
answers to the corresponding Hash $ H_i ,i = 0,1,2,3$ query.

\noindent\textbf{Setup.} At the beginning, challenger $C$ runs the
algorithm $ Setup(1^k )$ and acts as PKG. That is, he generates the
global public system parameters $(params,P_{Pub} )$ and the master
private key $s$. Then, he sends $(params,P_{Pub} )$ to the adversary
$A$.

\noindent\textbf{Probe.} We now describe how the challenger
simulates various queries.

\textbf{Simulator:} $ H_0 (ID_U )$

\noindent - If the record $ (ID_U ,Q_U ,S_U )$ is found in $L_0$,
then returns $Q_U$.

\noindent - Else chooses $Q_U$ randomly from $ G_1^* $; computes $
S_U  = sQ_U $; stores $ (ID_U ,Q_U ,$ $S_U )$ in $L_0$ and returns
$Q_U$.

\textbf{Simulator:} $H_1 (w)$

\noindent - Searches $ (w,h_1 ) $ in the list $L_1$. If such a pair
is found, returns $h_1$.

\noindent - Otherwise chooses $h_1$ randomly from $ \{ 0,1\} ^{n_2
}\times\{ 0,1\} ^{n_1 } \times G_1^*$, and puts $ (w,h_1 ) $ into
$L_1$ and returns $h_1$.

\textbf{Simulator:} $H_2 (m||ID_{A}||ID_{B})$

\noindent - Searches $(m||ID_{A}||ID_{B},h_2 )$ in List $L_2$. If
such a pair is found, returns $h_2$.

\noindent - Otherwise chooses $h_2$ randomly from $ Z_q^*$, and puts
$(m||ID_{A}||ID_{B},h_2 )$ into $L_2$ and returns $h_2$.

\textbf{Simulator:} $H_3 (m||X)$

\noindent - Searches $(m||X,h_3 )$ in the list $L_3$. If such a pair
is found, returns $h_3$.

\noindent - Otherwise chooses $h_3$ randomly from $ Z_q^*$, and puts
$(m||X,h_2 )$ into $L_3$ and returns $h_3$.

\textbf{Simulator:} $Extract(ID_U )$

We assume that $A$ makes the query $ H_0 (ID_U )$ before it makes
extract query for $ID_U$.

\noindent - Searches $L_0$ for the entry $ (ID_U ,Q_U ,S_U )$
corresponding to $ID_U$, and responds with $S_U$.

\textbf{Simulator:} $ Sign(ID_A ,m) $, $ Verify(ID_B ,\sigma )$

The challenger can easily answer these queries for the adversary.
Because the challenger initializes the system and he knows the
master key. So he can use signer $ID_A$'s private key to sign
message $m$ and use the receiver $ID_B$'s public key to verify the
signature $\sigma$ faithfully according to IDGSC-IN-SG. The only
difference is substituting the above Hash simulators for Hash
functions.

\textbf{Simulator:} $ Encrypt(ID_B ,m) $, $ Decrypt(ID_B
,\varepsilon )$

The challenger can get receiver $ID_B$'s public key and private key.
So he can supply these services for the adversary. Also the Hash
functions in the scheme use the above Hash simulators.

\textbf{Simulator:} $ GSC(ID_A ,ID_B ,m) $, $ GUC(ID_A ,ID_B
,\delta) $

The challenger can get sender $ID_A$'s public key and private key
and receiver $ID_B$'s public key and private key. So he can supply
these services for the adversary. Here, the Hash functions also use
the above Hash simulators.

\noindent\textbf{Exp 2}

In this experiment, we will remove the layer of encryption and
reduce the signcryption scheme to PSG scheme. In the Setup phase,
the challenger initializes the system just like he does in Exp 1. In
the Probe phase, besides following simulators, challenger acts same
with Exp 1.

\textbf{Simulator:} $ Sign(ID_A ,m) $, $ Verify(ID_B ,\sigma )$

Here, the challenger will follow PSG to accomplish these
simulations.

\textbf{Simulator:} $ GSC(ID_A ,ID_B ,m) $

Here, the challenger will keep another list $L_s$ to record the GSC
queries that the adversary asks.

\noindent - Selects $r$ uniformly from $Z_q^* $, and computes $ X =
rP $.

\noindent - Computes $ h_2  = H_2 (m||ID_A||ID_B) $ and $ h_3  = H_3
(m||X)$.

\noindent - Computes $ V = r^{ - 1} (h_2P + h_3 \cdot S_A )$.

\noindent - Selects $h_1$ uniformly from $ \{ 0,1\} ^{n_2 }  \times
\{ 0,1\} ^{n_1 }  \times G_1^*$ and adds $ (*,h_1 ) $ in List $L_1$.
The first element is vacant, and will be given some value later.

\noindent - Computes $ y = m||ID_A||V \oplus h_1 $ and adds $
(X,y,V,ID_A ,ID_B,m ) $ to List $L_s$.

\noindent - Outputs ciphertext $ (X,y)$. (Here, $ h_2 ,h_3$ come
from the corresponding Hash Simulators.)

\textbf{Simulator:} $ GUC(ID_A ,ID_B ,\delta) $

\noindent - Searches $ (*||ID_A||ID_B,* ) $ in the list $L_2$, if
such a record $ (m||ID_A||ID_B,h_2 ) $ is found, goes to the next
step. Else, returns $\bot$.

\noindent - Searches $ (m||*,* ) $ in the list $L_3$, if such a
record $ (m||X,h_3 ) $ is found, goes to the next step. Else,
returns $\bot$.

\noindent - Searches $ (X,*,*,ID_A ,ID_B,m ) $ in the list $L_s$, if
such a record $ (X,y,V,ID_A ,\\ID_B,m ) $ is found, goes to the next
step. Else, returns $\bot$.

\noindent - Checks that $ \hat e(X,V) = \hat e(P,P)^{h_2 } \cdot
\hat e(P_{Pub} ,Q_A )^{h_3}$, if not, returns $\bot$.

\noindent - Else computes $ w = \hat e(X,S_B )$ and $ h_1 = y \oplus
 m||ID_A||V$.

\noindent - Searches $ (*,h_1 ) $ in the list $L_1$, if such a
record is found, the first element defined to be $w$ and returns
$m$. Else, returns $\bot$.

Now we discuss the difference between Exp 1 and Exp 2. The adversary
can distinguish Exp 1 with Exp 2 if following events happened.
Firstly, during the Signcrypt query, if the adversary has made the
query $ H_1 (w) $, where $w$ happened to be the vacant value of some
record. The probability of such event happening is at most $ {q_1
}/{2^{n_4 } } $. The adversary made $q_s$ Signcrypt query. So the
probability of such events happening is at most $ {(q_1 \cdot q_s)
}/{2^{n_4 } } $ in total. Secondly, during the Unsigncrypt query, if
the adversary has guessed plaintext of some ciphertext. The
probability of such event happening is at most ${1}/{({2^{n_2 }
}\cdot{2^{n_1 } }\cdot{2^{n_3 } })} $. The adversary made $q_u$
Unsigncrypt query. So the probability of such events happening is at
most $ {{q_u }}/{({2^{n_2 } }\cdot{2^{n_1 } }\cdot{2^{n_3 } })} $ in
total. Let $S_1$ denote adversary can attack successfully in Exp 2.
So, we have:

$ |\Pr (S_0 ) - \Pr (S_1 )| \le {{(q_{h_1 }  \cdot q_s )}}/{{2^{n_4
} }} + {{q_u }}/{({2^{n_2 } }\cdot{2^{n_1 } }\cdot{2^{n_3 } })}$

\section{Proof of Theorem 4}

\textbf{Weak BCDH problem.} $(G_1 , + )$ and $(G_2 , \cdot )$ are
two cycle groups of prime order $q$, $P$ is a generator of $G_1$,  $
\hat e:G_1 \times G_1 \to G_2$ is a bilinear map between $G_1$ and
$G_2$. Given $ (P,aP,bP,cP,\frac{1}{c}P)$, where $a,b,c\in
Z_{q}^{*}$, the strong BDH problem is to compute $\hat
e(P,P)^{abc}$.

\textbf{Proof.} If there is an IND-CCA2 adversary $A$ of IDGSC in
the signcryption mode, then the challenger $C$ can use it to solve
the strong BDH problem. Let $ (P,aP,bP,cP,\frac{1}{c}P) $ be an
instance of the weak BCDH problem that $C$ wants to solve. At first,
$C$ runs the $ Setup(1^k )$ algorithm to produce parameters $params
$. It sets the public key as $ P_{Pub} = cP $, although it doesn't
know the master key $c$. And then $C$ sends $(params,P_{Pub} )$ to
the adversary $A$.

Besides the four lists $L_i ,i = 0,1,2,3 $, Challenger $C$ also
needs to keep  another list $L_s$ which are used to record answers
to the Signcrypt query.

\noindent\textbf{Phase 1}

\textbf{Simulator:} $ H_0 (ID_U )$

At the beginning, $C$ chooses $i_b$ uniformly at random from
${1,...q_0}$. We assume that $A$ doesn't make repeat queries.

\noindent - If $ i = i_b $ responds with $ H_0 (ID_U ) = bP $ and
sets $ ID_U   = ID_b$.

\noindent - Else chooses $k$ uniformly at random from $ Z_q^* $;
computes $ Q_U  = kP $ and $ S_U  = kP_{Pub} $; stores $ (ID_U ,Q_U
,$ $S_U,k )$ in $L_0$ and responds with $Q_U$.

\textbf{Simulator:} $H_1 (w)$

\noindent - Searches $ (w,h_1 ) $ in List $L_1$. If such a pair is
found, returns $h_1$.

\noindent - Otherwise chooses $h_1$ randomly from $ \{ 0,1\} ^{n_2 }
\times \{ 0,1\} ^{n_2 }  \times G_1^* $, and puts $ (w,h_1 ) $ into
$L_1$ and returns $h_1$.

\textbf{Simulator:} $H_2 (m||ID_{1}||ID_{2})$

\noindent - Searches $(m||ID_{1}||ID_{2},h_2 )$ in List $L_2$. If
such a pair is found, returns $h_2$.

\noindent - Otherwise chooses $h_2$ randomly from $ Z_q^*$, and puts
$(m||ID_{1}||ID_{2},h_2 )$ into $L_2$ and returns $h_2$.

\textbf{Simulator:} $H_3 (m||X)$

\noindent - Searches $(m||X,h_3 )$ in the list $L_3$. If such a pair
is found, returns $h_3$.

\noindent - Otherwise chooses $h_3$ randomly from $ Z_q^*$, and puts
$(m||X,h_3 )$ into $L_3$ and returns $h_3$.

\textbf{Simulator:} $Extract(ID_U )$

We assume that $A$ makes the query $ H_0 (ID_U )$ before it makes
extract query for $ID_U$.

\noindent - If $ID_U=ID_b$, aborts the simulation.

\noindent - Else, searches $L_0$ for the entry $ (ID_U ,Q_U ,S_U
,k)$ corresponding to $ID_U$, and responds with $S_U$.

\textbf{Simulator:} $ Sign(ID_1 ,m) $

We assume that $A$ makes the query $ H_0 (ID_1 )$ before $ Sign(ID_1
,m) $ query.

\noindent\textbf{Case 1:} $ ID_1  \ne ID_b$

\noindent - Find the entry $ (ID_1 ,Q_1 ,$ $S_1,k )$ in $L_0$.

\noindent - Selects $r$ uniformly from $Z_q^*$, and computes $X=rP$.

\noindent - Computes $ h_2  = H_2 (m||ID_1 ||0) $ and $ h_3  = H_2
(m||X) $.

\noindent - Computes $ V = r^{ - 1} (h_2 P  + h_3  \cdot S_1 ) $.

\noindent - Outputs $ (X,m||ID_1 ||V) $. (Here $ H_i $, $i=2,3$,
comes from the simulator above. )

\noindent\textbf{Case 2:} $ ID_1  = ID_b$

\noindent - Selects $r$ uniformly from $Z_q^*$, and computes
$X=rP_{Pub}$.

\noindent - Computes $ h_2  = H_2 (m||ID_1 ||0) $ and $ h_3  = H_3
(m||X) $.

\noindent - Computes $ V = r^{ - 1} (h_2 \cdot \frac{1}{c}P  + h_3
\cdot bP ) $.

\noindent - Outputs $ (X,m||ID_1 ||V) $. (Here $ H_i $, $i=2,3$,
comes from the simulator above. )

\textbf{Simulator:} $ Verify(ID_1 ,\sigma )$

\noindent - Computes $ h_2  = H_2 (m||ID_1 ||0) $, If $ ( m||ID_1
||0,h_2) \notin L_2$, returns $\bot$.

\noindent - Computes $ h_3  = H_3 (m||X) $, If $ ( m||X,h_{3})
\notin L_3 $, returns $\bot$.

\noindent - If $ID_1 \notin L_0$, returns $\bot$; else computes
$Q_0=H_{0}(ID_{1})$.

\noindent - Checks that $ \hat e(X,V) = \hat e(P,P )^{h_2 } \cdot
\hat e(P_{Pub} ,Q_1 )^{h_3 } $, if not, returns $\bot$. Else,
returns $\top$.

\textbf{Simulator:} $ Encrypt(ID_2 ,m )$

We assume that $A$ has made the $ H_0 (ID_2 )$ query before $
Encrypt(ID_2 ,m )$ query.

\noindent - Selects $r$ uniformly from $Z_q^* $, and computes $ X =
rP $.

\noindent - Computes $ h_2  = H_2 (m||0||ID_2) $ and $ h_3  = H_3
(m||X)$.

\noindent - Computes $ V = r^{ - 1} h_2P$.

\noindent - Computes $ Q_B  = H_0 (ID_2 )$ and $w = \hat e(P_{Pub}
,Q_2 )^{r}$.

\noindent - Computes $h_1  = H_1 (w) $ and $ y = m||0||V \oplus h_1
$.

\noindent - Outputs $ (X,y)$.(Here $H_i$, i=0,1,2,3, comes from the
simulator above.)

\textbf{Simulator:} $ Decrypt(ID_2 ,\varepsilon )$

We assume that $A$ makes the query $ H_0 (ID_2 )$ before $
Decrypt(ID_2 ,\varepsilon )$.

\noindent\textbf{Case 1:} $ ID_2  \ne ID_b$

\noindent - Find the entry $ (ID_2 ,Q_2 ,$ $S_2,k )$ in $L_0$.

\noindent - Computes $ w = \hat e(X,S_2 )$ and $ h_1 = H_1 (w)$.

\noindent - If $ (w,h_1 ) \notin L_1 $, returns $\bot$. Else,
computes $m||0||V = y \oplus h_1$.

\noindent - Computes $ h_2  = H_2 (m||0 ||ID_2) $, If $ (m||0
||ID_2,h_2) \notin L_2$, returns $\bot$.

\noindent - Computes $ h_3  = H_3 (m||X) $, If $ (m||X,h_{3}) \notin
L_3 $, returns $\bot$.

\noindent - Checks that $ \hat e(X,V) = \hat e(P,P )^{h_2 }$, if
not, returns $\bot$. Else, returns $m$.

\noindent\textbf{Case 2:} $ ID_2  = ID_b$

Step through the list $L_1$ with entries $(w,h_1)$ as follows:

\noindent - Computes $ m||0||V = y \oplus h_1$.

\noindent - If $m||0||ID_2  \in L_2 $, computes
$h_2=H_{2}(m||0||ID_2 )$; else moves to the next entry in $L_1$ and
begin again.

\noindent - If $m||X  \in L_3 $, computes $h_3=H_{3}(m||X )$; else
moves to the next entry in $L_1$ and begin again.

\noindent - Checks that $ \hat e(X,V) = \hat e(P,P )^{h_2 } $. If
so, returns $m$; else moves to the next entry in $L_1$ and begin
again.

\noindent - If no message has been returned after stepping through
$L_1$, return $\bot$.

\textbf{Simulator:} $ Signcrypt(ID_1 ,ID_2 ,m )$

We assume that $A$ makes the query $ H_0 (ID_1 )$ and $ H_0 (ID_2 )$
before making signcrypt query using identity $  ID_1 $ and $ ID_2 $.

\noindent\textbf{Case 1:} $ ID_1  \ne ID_b$

\noindent - Find the entry $ (ID_1 ,Q_1 ,$ $S_1,k )$ in $L_0$.

\noindent - Selects $r$ uniformly from $Z_q^* $, and computes $ X =
rP $.

\noindent - Computes $ h_2  = H_2 (m||ID_1||ID_2) $ and $ h_3  = H_3
(m||X)$.

\noindent - Computes $ V = r^{ - 1} (h_2P+h_{3}S_1)$.

\noindent - Computes $ Q_2  = H_0 (ID_2 )$ and $w = \hat e(P_{Pub}
,Q_2 )^{r }$.

\noindent - Computes $h_1  = H_1 (w) $ and $ y = m||ID_1||V \oplus
h_1 $.

\noindent - Outputs $ (X,y)$.(Here $H_i$, i=0,1,2,3, comes from the
simulator above.)

\noindent\textbf{Case 2:} $ ID_1  = ID_b$

\noindent - Find the entry $ (ID_2 ,Q_2 ,$ $S_2,k )$ in $L_0$.

\noindent - Selects $r$ uniformly from $Z_q^*$, and computes
$X=rP_{Pub}$.

\noindent - Computes $ h_2  = H_2 (m||ID_1 ||ID_2) $ and $ h_3  =
H_3 (m||X) $.

\noindent - Computes $ V = r^{ - 1} (h_2\cdot\frac{1}{c} P  + h_3
\cdot bP ) $.

\noindent - Computes $w = \hat e(X ,S_2 )$, $h_1  = H_1 (w) $ and $
y = m||ID_1||V \oplus h_1 $.

\noindent - Outputs $ (X,y)$.(Here $H_i$, i=1,2,3, comes from the
simulator above.)

\textbf{Simulator:} $ Unsigncrypt(ID_1 ,ID_2 ,\varepsilon )$

We assume that $A$ makes the query $ H_0 (ID_1 )$ and $ H_0 (ID_2 )$
before making this query using these identities.

\noindent\textbf{Case 1:} $ ID_2  \ne ID_b$

\noindent - Find the entry $ (ID_2 ,Q_2 ,$ $S_2,k )$ in $L_0$.

\noindent - Computes $ w = \hat e(X,S_2 )$ and $ h_1 = H_1 (w)$.

\noindent - If $ (w,h_1 ) \notin L_1 $, returns $\bot$. Else,
computes $m||ID_1||V = y \oplus h_1$.

\noindent - Computes $ h_2  = H_2 (m||ID_1||ID_2) $, If $ (m||ID_1
||ID_2,h_2) \notin L_2$, returns $\bot$.

\noindent - Computes $ h_3  = H_3 (m||X) $, If $ (m||X,h_{3}) \notin
L_3 $, returns $\bot$.

\noindent - If $ID_1  = ID_2 $ or $ID_1 \notin L_0$, returns $\bot$;
else computes $ Q_1  = H_0 (ID_1 ) $.

\noindent - Checks that $  \hat e(X,V) = \hat e(P,P)^{h_2 } \cdot
\hat e(P_{Pub} ,Q_1 )^{h_3 } $, if not, returns $\bot$. Else,
returns $m$.

\noindent\textbf{Case 2:} $ ID_2  = ID_b$

Step through the list $L_1$ with entries $(w,h_1)$ as follows:

\noindent - Computes $ m||ID_1||V = y \oplus h_1$.

\noindent - If $ID_1  = ID_2 $ or $ID_1 \notin L_0$, moves to the
next entry in $L_1$ and begin again; else computes $ Q_1  = H_0
(ID_1 ) $.

\noindent - If $m||ID_1||ID_2  \in L_2 $, computes
$h_2=H_{2}(m||ID_1||ID_2 )$; else moves to the next entry in $L_1$
and begin again.

\noindent - If $m||X  \in L_3 $, computes $h_3=H_{3}(m||X )$; else
moves to the next entry in $L_1$ and begin again.

\noindent - Checks that $  \hat e(X,V) = \hat e(P,P)^{h_2 } \cdot
\hat e(P_{Pub} ,Q_1 )^{h_3 } $. If so, returns $m$; else moves to
the next entry in $L_1$ and begin again.

\noindent - If no message has been returned after stepping through
$L_1$, return $\bot$.

\textbf{Challenge.} At the end of Phase 1, the adversary $A$ outputs
two identities, $ID_A$ and $ID_B$, two messages, $m_1$ and $m_2$. If
$ID_B  \ne ID_b$, aborts the simulation; else it sets $ X^*  = aP $
and then chooses $ \gamma  \in \{ 0,1\} $, and $ y^*  \in \{ 0,1\}
^{n_2 }  \times \{ 0,1\} ^{n_2 }  \times G_1^* $ at random. At last,
it returns the challenge ciphertext $ \delta ^*  = (X^* ,y^* ) $ to
$A$.

\textbf{Phase 2.}

The queries made by in Phase 2 are responded in the same way as
those made by in Phase 1. Here, the queries follow the restrictions
that are defined in Game 6.

\textbf{Guess.}

At the end of Phase 2, $A$ outputs a bit $\gamma^{'}$. If
$\gamma^{'}=\gamma$, the challenger $C$ outputs the answer to the
weak BCDH problem:

$w^*  = \hat e(X^* ,S_B ) = \hat e(P,P)^{abc}$

Let's analyze the probability that the simulation can succeed. There
are two simulators need to be noted. First, in the challenge stage,
the simulator hopes that the adversary chosen $ID_b$ as the target
recipient identity. This will be the case with probability at least
$ {1}/{q_{0 }} $. If this is not the case, there will be an error
when the adversary tried to make query $ Extract(ID_b ) $. Second,
in Phase 2, if the adversary makes query $ H_1 (w = \hat
e(P,P)^{abc} ) $, the simulation will fail. However, with
probability $ {1}/{q_{1 }} $ the challenger can guess the answer of
weak BCDH problem from the records in List $L_1$. From the above
remarks we conclude that the challenger can solve the weak BCDH
problem with probability at least: $  Adv_{A^{idgsc - in - sc}
}^{ind - cca2} (t,p) /{{(q_{0 \cdot } q_1) }}$.











\end{document}